\documentclass[12pt]{article}
\usepackage{epsfig}
\usepackage{graphicx}
\def\br(#1,#2){\left\langle#1#2\right\rangle}
\def\sq(#1,#2){\left[#1#2\right]}
\def\s(#1,#2){s_{#1 #2}}
\def\t(#1,#2,#3){s_{#1 #2 #3}}

\begin{document}
\begin{titlepage}

\hspace*{\fill}\parbox[t]{5.5cm}
{hep-ph/0510362\\
CERN-PH-TH/2005-064\\
FERMILAB-PUB-05/487-T\\
ILL-(TH)-05-04\\
\today} \vskip2cm
\begin{center}
{\Large \bf Production of a $Z$ Boson and Two Jets \\
\bigskip  with One Heavy-Quark Tag} \\
\medskip
\bigskip\bigskip\bigskip\bigskip
{\large  {\bf J.~Campbell}$^1$,
         {\bf R.~K.~Ellis}$^{1,2}$,
         {\bf F.~Maltoni}$^{1,3}$,
     and {\bf S.~Willenbrock}$^{4}$} \\
\bigskip\bigskip\medskip
$^{1}$CERN, CH-1211 Geneva 23, Switzerland\\ \bigskip
$^{2}$Theoretical Physics Department, Fermi National Accelerator Laboratory \\
P.~O.~Box 500, Batavia, IL\ \ 60510 \\ \bigskip
$^{3}$Institut de Physique Th\'eorique, Universit\'e Catholique de Louvain \\
Chemin du Cyclotron, 2, B-1348 Louvain-la-Neuve, Belgium \\
\bigskip
$^{4}$Department of Physics, University of Illinois at Urbana-Champaign \\
1110 West Green Street, Urbana, IL\ \ 61801 \\ \bigskip
\end{center}

\bigskip\bigskip\bigskip

\begin{abstract}
We present a next-to-leading-order calculation of the production of
a $Z$ boson with two jets, one or more of which contains a heavy
quark ($Q=c,b$).  We show that the cross section with only one
heavy-quark jet is larger than that with two heavy-quark jets at
both the Fermilab Tevatron and the CERN LHC.  These processes are
the dominant irreducible backgrounds to a Higgs boson produced in
association with a $Z$ boson, followed by $h\to b\bar b$. Our
calculation makes use of a heavy-quark distribution function, which
resums collinear logarithms and makes the next-to-leading-order
calculation tractable.
\end{abstract}

\end{titlepage}

\section{Introduction}
\label{sec:intro}

The discovery of new physics at hadron colliders often relies on a
detailed understanding of standard-model background processes.
Prominent among these is the production of weak bosons ($W,Z$) in
association with jets, one or more of which contains a heavy quark
($Q=c,b$).  The prime example is the discovery of the top quark at
the Fermilab Tevatron, which required a thorough understanding of
the $W+$jets background, with one or more heavy-quark jets
\cite{Abe:1995hr,Abachi:1995iq,Acosta:2001ct}. The discovery of
single-top-quark production via the weak interaction requires an
even more sophisticated understanding of this background
\cite{Abazov:2005zz,Acosta:2004bs}.

In this paper we present a next-to-leading-order (NLO) calculation
of the production of a $Z$ boson in association with two or more
jets, one or more of which contains a heavy quark. This is the
dominant irreducible background to the production of a Higgs boson
in association with a $Z$ boson, followed by $h\to b\bar b$
\cite{Stange:1993ya,Stange:1994bb}.  This signal is currently being
sought at the Tevatron with either one or both $b$ jets tagged
\cite{Acosta:2005ga}.  The calculation we present provides the
irreducible background at NLO for both cases.

The production of a $Z$ boson plus two jets with at least one
heavy-quark jet is also a background to the production of a Higgs
boson in association with one or more $b$ jets, which is a discovery
mode for a supersymmetric Higgs boson at large values of $\tan\beta$
\cite{Dicus:1988cx,Kunszt:1991qe,Kao:1995gx,Dai:1994vu,Dai:1996rn,
Richter-Was:1997gi,Barger:1997pp,Choudhury:1998kr,Huang:1998vu,Drees:1997sh,
Carena:1998gk,Diaz-Cruz:1998qc,Balazs:1998nt,Balazs:1998sb,Dicus:1998hs,
unknown:1999fr,Carena:2000yx,Affolder:2000rg,Campbell:2002zm,Maltoni:2003pn,
Harlander:2003ai,Dittmaier:2003ej,Dawson:2003kb,Dawson:2004sh,
Campbell:2004pu}.\footnote{The minimal supersymmetric standard model
requires two Higgs doublets; the ratio of their vacuum expectation
values is $\tan\beta\equiv v_2/v_1$.}  This background process is
also a benchmark for this Higgs discovery channel. The search for a
supersymmetric Higgs boson via this mode is underway at the Tevatron
\cite{Affolder:2000rg,Acosta:2005bk,Abazov:2005yr}, and will be
vigorously pursued at the CERN Large Hadron Collider (LHC)
\cite{Abdullin:2005yn,Schumacher:2004da}.

\begin{figure}[ht]
\begin{center}
\vspace*{.2cm}
\hspace*{0cm}
\epsfxsize=10cm \epsfbox{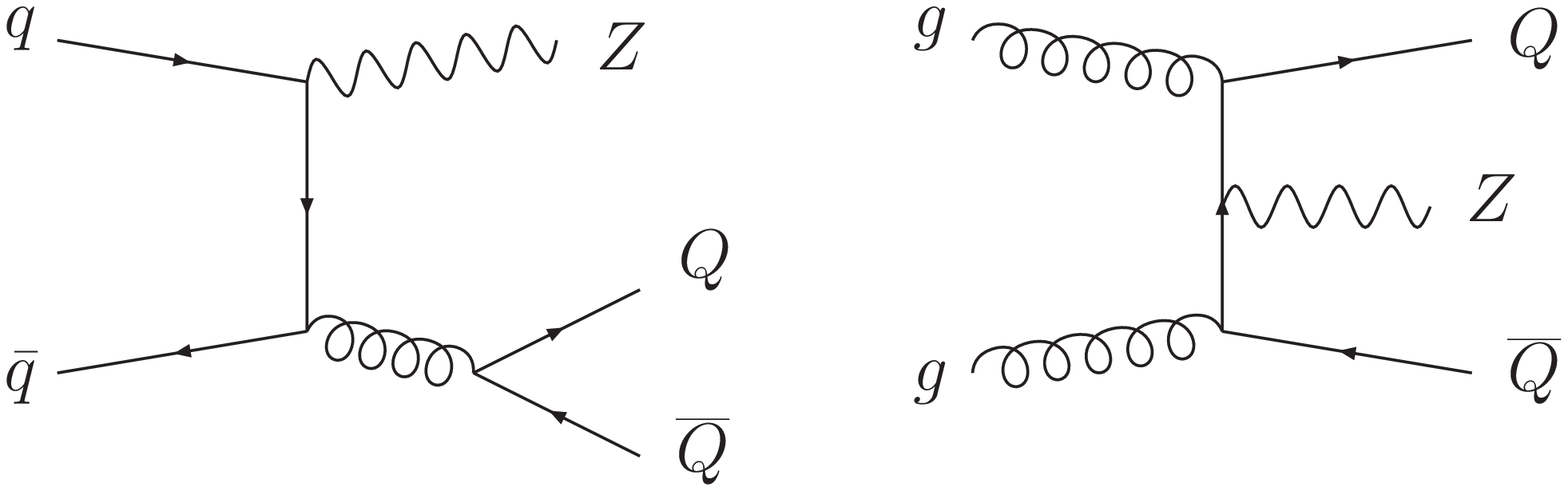}
\vspace*{-.8cm}
\end{center}
\caption{Diagrams contributing to the associated production of a $Z$
boson and two high-$p_T$ heavy quarks ($Q=c,b$).} \label{fig:ZQQ}
\end{figure}

\begin{figure}[ht]
\begin{center}
\vspace*{0cm}
\hspace*{0cm}
\epsfxsize=10cm \epsfbox{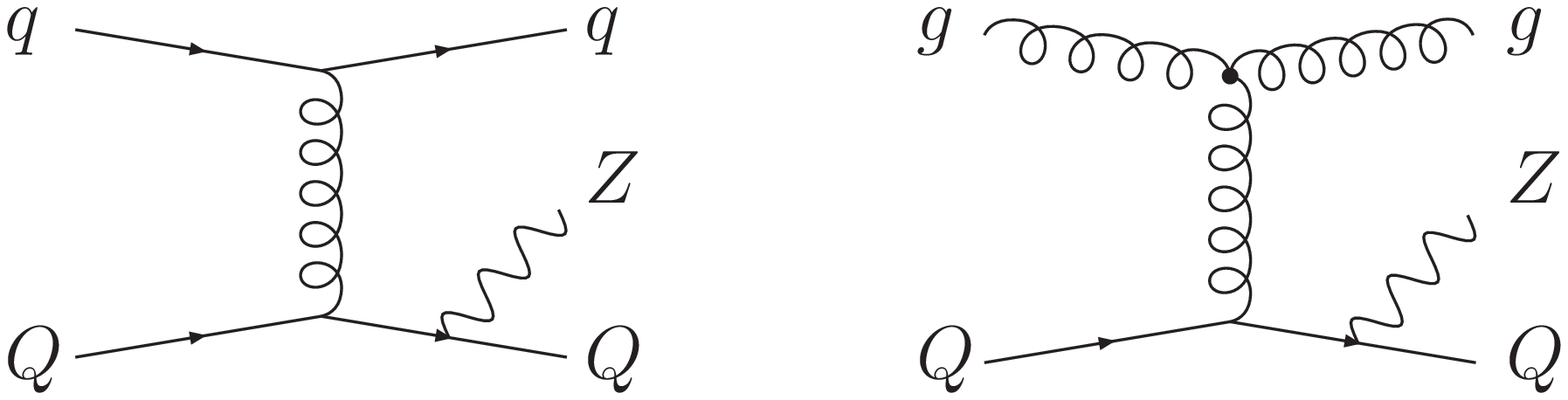}
\end{center}
\caption{Diagrams contributing to the associated production of a $Z$
boson and two high-$p_T$ jets, one of which contains a heavy quark
($Q=c,b$).} \label{fig:ZQj}
\end{figure}

When one considers the production of a heavy quark at a hadron
collider, ones first thought is usually of a virtual gluon splitting
into a final-state $Q\overline Q$ pair, as shown in
Fig.~\ref{fig:ZQQ}(a), or via gluon fusion, as shown in
Fig.~\ref{fig:ZQQ}(b). However, initial gluons splitting into a
$Q\overline Q$ pair is just as important a source at the Tevatron,
and even more important at the LHC, when only one heavy quark is
observed at high transverse momentum ($p_T$).  In that situation, it
is advantageous to think of the initial gluon as splitting into a
collinear $Q\overline Q$ pair, with one heavy quark remaining at low
$p_T$ while the other heavy quark participates in the hard
scattering and emerges at high $p_T$. The heavy quark that
participates in the hard scattering can be treated as part of the
proton sea, with a parton distribution function that is calculated
perturbatively from the Dokshitzer-Gribov-Lipatov-Altarelli-Parisi
(DGLAP) evolution equations \cite{Aivazis:1993pi,Collins:1998rz}.
The production of a $Z$ boson with two jets, one of which contains a
heavy quark, then proceeds as shown in Fig.~\ref{fig:ZQj}.

The reasons for using a heavy-quark distribution function are
twofold.  First, it resums collinear logarithms of the form $\ln
Q/m_Q$ to all orders, where $Q$ is the scale of the hard
scattering.  Second, it simplifies the leading-order process,
which makes a higher-order calculation tractable.

This paper completes the NLO calculation of $Z+$jets, with one or
more heavy quarks, up to two jets \cite{Campbell:2000bg}.  The
production of a $Z$ plus one jet, with one or more heavy quarks, was
presented in Ref.~\cite{Campbell:2003dd}, and agrees well with data
from the Tevatron \cite{Abazov:2004zd}.  The inclusive production of
a $Z$ with one or more heavy quarks was presented in
Ref.~\cite{Maltoni:2005wd}.

\section{$ZQj$ at NLO}\label{sec:ZQj}

The leading-order (LO) processes for $Z$ plus two jets, one or more
of which contains a heavy quark, are $q\bar q(gg)\to ZQ\overline Q$
(Fig.~\ref{fig:ZQQ}) and $Qq(g)\to ZQq(g)$ (Fig.~\ref{fig:ZQj}). The
LO cross sections are given in Table~\ref{tab:lo}, with the jets
satisfying the conditions $p_T>15$ GeV, $|\eta|<2$ (2.5 at the LHC),
and $\Delta R_{jj}>0.7$. At the Tevatron, the leading-order cross
sections for these two classes of processes are comparable; at the
LHC, the latter dominates. The processes $Qq(g)\to ZQq(g)$ are
relatively more important at the LHC than at the Tevatron because
they are initiated by a heavy sea quark, whose distribution function
rises at small values of $x$. Furthermore, $Qg\to ZQg$ is larger
than $Qq\to ZQq$ at the LHC, while they are comparable at the
Tevatron, since the former involves the gluon distribution function,
which is large at small values of $x$. For the same reason, $gg\to
ZQ\overline Q$ is dominant at the LHC compared with $q\bar q\to
ZQ\overline Q$, while the latter is more important at the Tevatron.


\begin{table}[p]
\caption{Leading-order cross sections (pb) for $Z$ boson plus two
jets, one or two of which contains a heavy quark, at the Tevatron
($\sqrt{s}=1.96$ TeV $p\bar p$) and the LHC ($\sqrt{s}=14$ TeV
$pp$). A jet lies in the range $p_T>15$ GeV and $|\eta|<2$ (2.5 at
the LHC), with $\Delta R_{jj}>0.7$.  No branching ratios or tagging
efficiencies are included. The labels on the columns have the
following meaning: $ZQj=$ exactly two jets, one of which contains a
heavy quark; $ZQ\overline Q=$ exactly two jets, both of which
contain a heavy quark. The CTEQ6L1 parton distribution functions are
used \cite{Pumplin:2002vw}, with the factorization and
renormalization scales chosen as $\mu_F=\mu_R=M_Z$.  Also given, in
square brackets, is the LO cross section obtained without using a
heavy-quark distribution function, from $gq(g)\to ZQ\overline
Qq(g)$.} \addtolength{\arraycolsep}{0.2cm}
\renewcommand{\arraystretch}{1.5}
\medskip
\begin{center}
\hspace*{-1cm}
\begin{tabular}[5]{|c|cc|cc|}
\hline \hline
& \multicolumn{4}{|c|}{$\sigma$ (pb)}\\
\cline{2-5}
Process & \multicolumn{2}{|c|}{Tevatron} & \multicolumn{2}{c|}{LHC} \\
\cline{2-5}
           & $ZQj$           &  $ZQ\overline Q$ &   $ZQj$    & $ZQ\overline Q$   \\
\hline
 $bq\to Zbq+bg\to Zbg$   & 0.89+1.29=2.18 [1.91]   &  --  &  76+276=352 [191]   &  --   \\
 $q\bar q\to Zb\bar b+gg\to Zb\bar b$   & --  &  1.89+0.58=2.47  &  --  &  13+96=109   \\
\hline
 $cq\to Zcq+cg\to Zcg$   & 1.37+1.83=3.20 [3.26]   &  --  &  98+345=443 [271]  &  --   \\
 $q\bar q\to Zc\bar c+gg\to Zc\bar c$   & --  &  1.89+0.45=2.34  &  --  &  12+75=87   \\
\hline\hline
\end{tabular}
\end{center}
\label{tab:lo}
\end{table}

As is often the case, the distinction between the various
processes is lost once one goes beyond leading order.
Fig.~\ref{fig:ZQQj}(a) shows a Feynman diagram which ostensibly
contributes to the NLO correction to $q\bar q\to ZQ\overline Q$,
while the Feynman diagram in Fig.~\ref{fig:ZQQj}(b) ostensibly
contributes to the NLO correction to $Qq\to ZQq$, and the diagram
in Fig.~\ref{fig:ZQQj}(c) to the NLO correction to $gg\to
ZQ\overline Q$. However, all three diagrams contribute to the same
amplitude, and therefore interfere.  Thus one cannot uniquely
identify them with any of the leading-order processes.

\begin{figure}[ht]
\begin{center}
\vspace*{0cm} \hspace*{0cm} \epsfxsize=16cm \epsfbox{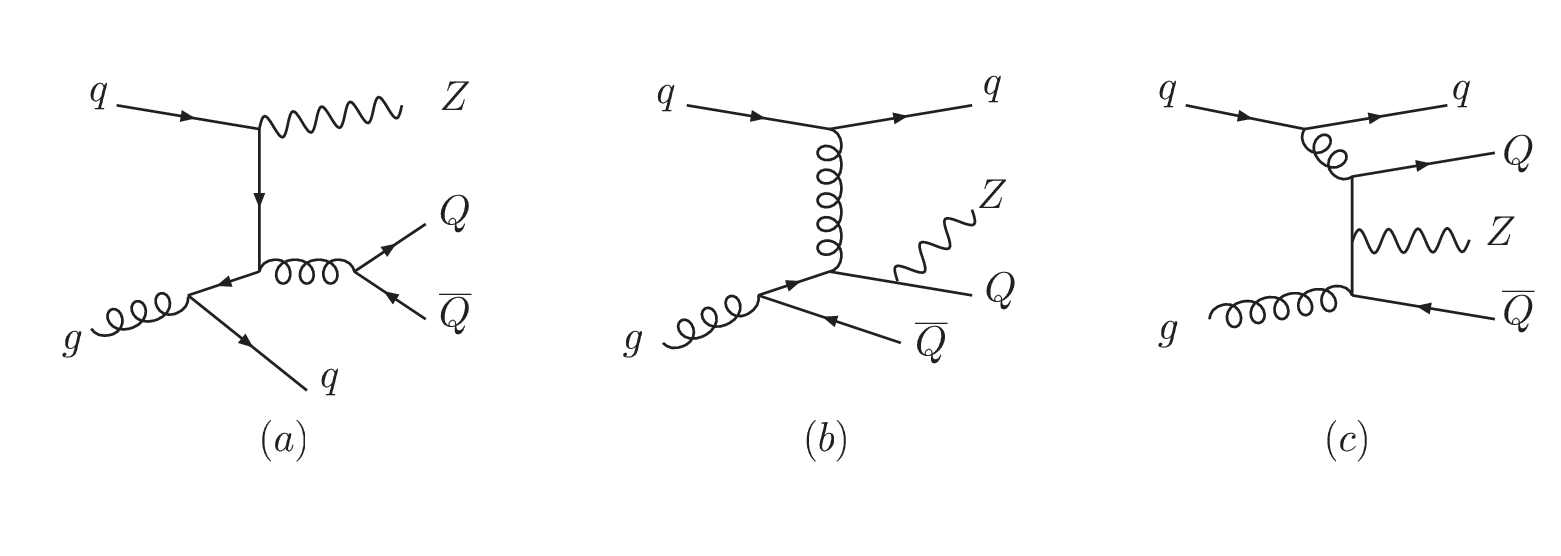}
\vspace*{-.8cm}
\end{center}
\caption{Diagrams contributing to the NLO correction to the
associated production of a $Z$ boson and two high-$p_T$ jets, one or
more of which contains a heavy quark ($Q=c,b$).} \label{fig:ZQQj}
\end{figure}

The next-to-leading-order calculations in this paper were
performed with the Monte-Carlo code MCFM \cite{Ellis:1999ec}.  The
leading-order calculations were performed both with this code and
with MadEvent \cite{Maltoni:2002qb}.  The NLO corrections are
included in the code MCFM by implementing the virtual helicity
amplitudes of Ref.~\cite{Bern:1997sc}. These are given in the
four-dimensional helicity scheme, which is used throughout the
calculation. The real corrections are adapted from
Ref.~\cite{Nagy:1998bb} and singularities are handled using the
dipole subtraction method~\cite{Catani:1996vz}. Since the matrix
elements contain the decay of the gauge boson into massless
particles, the code is general enough to provide results for final
states from $\gamma^*,Z^*\to \ell^+\ell^-$ (including
interference). For this paper we specialize to the case of a real
$Z$ boson.

The processes involved in the calculation are as follows:
\begin{itemize}
\item $q\bar q\to ZQ\overline Q$ at tree level (Fig.~\ref{fig:ZQQ}) and one loop
\item $gg\to ZQ\overline Q$ at tree level (Fig.~\ref{fig:ZQQ}) and one loop
\item $Qq\to ZQq$ at tree level (Fig.~\ref{fig:ZQj}) and one loop
\item $Qg\to ZQg$ at tree level (Fig.~\ref{fig:ZQj}) and one loop
\item $q\bar q\to ZQ\overline Qg$ at tree level
\item $gg\to ZQ\overline Qg$ at tree level
\item $Qg\to ZQgg$ at tree level
\item $Qq\to ZQqg$ at tree level
\item $gq\to ZQ\overline Qq$ at tree level (Fig.~\ref{fig:ZQQj})
\item $Qg\to ZQq\bar q$ at tree level
\end{itemize}
We also include the processes $QQ'\to ZQQ'$, $Q\overline Q'\to
ZQ\overline Q'$, and $Q\overline Q\to ZQ'\overline Q'$ at LO.
These processes are already small at LO, so it is safe to neglect
their NLO corrections.  For charm final states, we only take
$Q=Q'=c$; for bottom, $Q,Q'=c,b$.

The results of our calculation are presented in
Tables~\ref{tab:tev_excl} and \ref{tab:lhc_excl}. The columns
$ZQj$ and $ZQ\overline Q$ contain the leading-order cross sections
(in parentheses) for the processes $Qq(g)\to ZQq(g)$ and $q\bar
q(gg)\to ZQ\overline Q$, respectively, taken from
Table~\ref{tab:lo}. We require the jets to satisfy the conditions
$p_T>15$ GeV, $|\eta|<2$ (2.5 at the LHC), and $\Delta
R_{jj}>0.7$. The heavy-quark mass is neglected here and throughout
the calculation (except where noted).  To obtain the NLO cross
sections for $ZQj$ and $ZQ\overline Q$, which involve the
radiation of additional partons ({\it e.g.}, Fig.~\ref{fig:ZQQj}),
two partons are combined into a single jet by adding their
four-momenta if $\Delta R_{jj}<0.7$. If the combined partons are
both heavy quarks, then the process contributes to the column
labeled $Z(Q\overline Q)j$. This is a $Z+2j$ event in which one
jet contains two heavy quarks, which changes the tagging
probability for that jet \cite{Acosta:2004nj}. It is calculated
with a finite heavy-quark mass ($m_c=1.4$ GeV, $m_b=4.75$ GeV) in
order to regulate the logarithmic divergence present when the
heavy quarks are collinear. If all three partons are well
separated, then the process contributes to either $ZQ\overline Qj$
or $ZQjj$.\footnote{The matrix elements for both $Z(Q\overline
Q)j$ and $ZQ\overline Qj$ are calculated with a finite heavy-quark
mass, but massive phase space is used only in the former.  We
checked that the quark mass is numerically irrelevant to the
latter process.}


\begin{table}[t]
\caption{Cross sections (pb) for $Z$-boson plus two (or more) jets,
one or more of which contains a heavy quark, at the Tevatron
($\sqrt{s}=1.96$ TeV $p\bar p$). A jet lies in the range $p_T>15$
GeV and $|\eta|<2$. Two final-state partons are merged into a single
jet if $\Delta R_{jj}<0.7$. No branching ratios or tagging
efficiencies are included. Numbers in parenthesis are leading-order
results. The labels on the columns have the following meaning:
$ZQj=$ exactly two jets, one of which contains a heavy quark;
$ZQ\overline Q=$ exactly two jets, both of which contain a heavy
quark; $Z(Q\overline Q)j=$ exactly two jets, one of which contains a
heavy-quark pair; $ZQ\overline Qj=$ exactly three jets, two of which
contain a heavy quark; $ZQjj=$ exactly three jets, one of which
contains a heavy quark.  For the last set of processes, the labels
mean: $Zjj=$ exactly two jets, including heavy quarks; $Zjjj=$
exactly three jets, including heavy quarks. For $ZQj$ and
$ZQ\overline Q$, both the leading-order (in parentheses) and
next-to-leading-order cross sections are given. The CTEQ6M parton
distribution functions are used throughout, except for the LO cross
sections in parentheses, where CTEQ6L1 is used
\cite{Pumplin:2002vw}. The factorization and renormalization scales
are chosen as $\mu_F=\mu_R=M_Z$.}

\addtolength{\arraycolsep}{0.2cm}
\renewcommand{\arraystretch}{1.5}
\medskip

\begin{center}
\begin{tabular}[5]{|c|ccccc|}
\hline \hline \multicolumn{1}{|c|}{Tevatron} &
\multicolumn{5}{c|}{$\sigma$ (pb)}\\[1pt]
\hline
         &  $ZQj$  & $ZQ\overline Q$ & $Z(Q\overline Q)j$  &   $ZQ\overline Qj$  &  $ZQjj$ \\
\hline
bottom   & (2.18) 5.23  & (2.47) 3.07  &  0.634 &  0.672 & 0.326 \\
charm    & (3.20) 7.49  & (2.34) 2.75  &  2.00  &  0.621 & 0.495 \\
\hline & \multicolumn{3}{|c}{$Zjj$} & \multicolumn{2}{c|}{$Zjjj$} \\
\hline $Z+$jets & \multicolumn{3}{|c}{(163) 182} & \multicolumn{2}{c|}{22.9} \\
\hline
\end{tabular}
\end{center}

\label{tab:tev_excl}
\end{table}


\begin{table}[t]
\caption{Same as Table~\ref{tab:tev_excl}, except at the LHC
($\sqrt{s}=14$ TeV $pp$).  A jet lies in the range $p_T>15$ GeV
and $|\eta|<2.5$.}

\addtolength{\arraycolsep}{0.2cm}
\renewcommand{\arraystretch}{1.5}
\medskip
\begin{center}
\begin{tabular}[5]{|c|ccccc|}
\hline \hline \multicolumn{1}{|c|}{LHC} &
\multicolumn{5}{c|}{$\sigma$ (pb)}\\[1pt]
\hline
         &  $ZQj$  & $ZQ\overline Q$ & $Z(Q\overline Q)j$  &   $ZQ\overline Qj$  &  $ZQjj$ \\
\hline
bottom   & (352) 421 & (109)  92.1  &  23.5 &  60.8 & 92.1 \\
charm    & (443) 623 & (87) 75.2  &  58.6 &  49.3 & 123 \\
\hline & \multicolumn{3}{|c}{$Zjj$} & \multicolumn{2}{c|}{$Zjjj$} \\
\hline $Z+$jets & \multicolumn{3}{|c}{(6090) 4840} & \multicolumn{2}{c|}{1810} \\
\hline
\end{tabular}
\end{center}

\label{tab:lhc_excl}
\end{table}


We checked that the effect of the heavy-quark mass is negligible
by comparing $ZQ\overline Q$ at tree level with and without a
finite quark mass.  Similarly, we found that the heavy-quark mass
is also negligible for $ZQ\overline Qj$ at tree level, as
expected.

The NLO correction to $ZQj$ is quite sizable at the Tevatron, more
than 100\%.  One of the reasons is that the process $q\bar q\to
ZQ\overline Qg$ (where one of the heavy quarks is outside the
acceptance) makes a relatively large contribution.  This is not
really a NLO correction to $Qq(g)\to ZQq(g)$, but rather a new
channel. This contribution is about 1.2 pb for both bottom and
charm at the Tevatron.  In contrast, this process makes a
relatively small contribution to $ZQj$ at the LHC, only 10 pb for
bottom and 4 pb for charm.

We also list in Tables~\ref{tab:tev_excl} and \ref{tab:lhc_excl}
the LO (in parentheses) and NLO cross sections for $Zjj$
\cite{Campbell:2000bg,Campbell:2002tg,Campbell:2003hd}, and the LO
cross section for $Zjjj$.  In these cross sections we have
included the contribution from light partons as well as heavy
quarks.  Thus, for example, the fraction of $Z+2j$ events in which
only one of the jets contains heavy quarks is given by
$[ZQj+Z(Q\overline Q)j]/Zjj$.

In Tables~\ref{tab:tev_incl} and \ref{tab:lhc_incl} we list the
results in an inclusive manner, and also include the NLO result
for $Z$ plus one heavy-quark jet \cite{Campbell:2003dd}. In
addition, we give the NLO cross section for $Zj$
\cite{Campbell:2003dd}, including both light and heavy partons.
Thus $ZQ+X$ is the cross section for a $Z$ plus at least one
heavy-quark jet; it is the sum of the columns labeled $ZQ$, $ZQj$,
and $ZQ\overline Q$ in Tables 1 and 2 of
Ref.~\cite{Campbell:2003dd}, including all contributing processes.
$Z(Q\overline Q)$ contains one jet which contains a heavy-quark
pair. It is calculated at LO, including the heavy-quark mass in
order to regulate the collinear divergence present in $q\bar q\to
Z(Q\overline Q)$. $ZQj+X$ has at least two jets, one of which
contains a heavy quark; it is the sum of the columns labeled $ZQj$
and $ZQjj$ in Tables~\ref{tab:tev_excl} and \ref{tab:lhc_excl}.
$ZQ\overline Q+X$ has at least two jets, two of which contain a
heavy quark; it is the sum of the columns labeled $ZQ\overline Q$
and $ZQ\overline Qj$ in Tables~\ref{tab:tev_excl} and
\ref{tab:lhc_excl}.  Finally, $Z(Q\overline Q)j$ contains two
jets, one of which contains a heavy-quark pair; it is calculated
at LO with a finite quark mass.

We estimate the uncertainties in the NLO inclusive cross sections
by varying the renormalization scale, the factorization scale, and
the parton distribution functions independently.  The
renormalization scale is varied over $\mu_R = (0.5 - 2) M_Z$, with
$\mu_F = M_Z$. Similarly, we vary $\mu_F = (0.5 - 2) M_Z$ with
$\mu_R = M_Z$. The uncertainties due to scale variation are
significantly reduced at NLO in comparison with the LO
calculation. Significant renormalization-scale dependence remains
since the process is ${\cal O}(\alpha_S^2)$.  The parton
distribution functions are varied over the 41 different sets
contained in CTEQ6M \cite{Pumplin:2002vw,Huston:2005jm}. 
A symmetric uncertainty
on the cross sections can be obtained from these sets by using
Eq.~(26) of Ref.~\cite{Pumplin:2001ct}. In addition, we have
studied the deviation of the cross section in each direction
separately, as in Ref.~\cite{Nadolsky:2001yg}. However, we find
that the difference between the positive and negative deviations
is not significant in any of the cases and thus we report only
symmetric uncertainties.

As an example of the use of these tables, consider the inclusive
cross section for $Z+$ 2 jets with one heavy-quark tag.  This
cross section is obtained from
\begin{equation}
\sigma = \epsilon_Q \sigma_{ZQj+X} + 2 \epsilon_Q (1-\epsilon_Q)
\sigma_{ZQ\overline Q+X}
        + \epsilon_{Q\overline Q} \sigma_{Z(Q\overline Q)j}
\end{equation}
where $\epsilon_Q$ is the tagging probability for a heavy-quark jet,
and $\epsilon_{Q\overline Q}$ is the tagging probability for a jet
containing a heavy-quark pair.


\begin{table}[p]
\caption{Inclusive cross sections (pb) for $Z$ boson plus one or two
jets, one or two of which contains a heavy quark, at the Tevatron
($\sqrt{s}=1.96$ TeV $p\bar p$). A jet lies in the range $p_T>15$
GeV and $|\eta|<2$. Two final-state partons are merged into a single
jet if $\Delta R_{jj}<0.7$. No branching ratios or tagging
efficiencies are included. Numbers in parenthesis are leading-order
results. The labels on the columns have the following meaning:
$ZQ+X=$ at least one jet, at least one of which contains a heavy
quark; $Z(Q\overline Q)=$ one jet which contains a heavy-quark pair;
$ZQj+X=$ at least two jets, one of which contains a heavy quark;
$ZQ\overline Q+X=$ at least two jets, two of which contain a heavy
quark; $Z(Q\overline Q)j=$ two jets, one of which contains a
heavy-quark pair.  The last row gives the inclusive cross section
for jets containing both light and heavy partons. The CTEQ6M parton
distribution functions are used throughout, except for the LO cross
sections in parentheses, where CTEQ6L1 is used
\cite{Pumplin:2002vw}.  The factorization and renormalization scales
are chosen as $\mu_F=\mu_R=M_Z$.  The uncertainties are from the
variation of the renormalization scale, the factorization scale, and
the parton distribution functions, respectively.}

\addtolength{\arraycolsep}{0.2cm}
\renewcommand{\arraystretch}{1.5}
\medskip
\begin{center}
\hspace*{-1.2cm}
\begin{tabular}[5]{|c|cc|ccc|}
\hline \hline \multicolumn{1}{|c|}{Tevatron} &
\multicolumn{5}{c|}{$\sigma$ (pb)}\\[1pt]
\hline
& \multicolumn{2}{|c|}{$Z+1$ jet $+X$} & \multicolumn{3}{c|}{$Z+2$ jets $+X$} \\
\cline{2-6}
           & $ZQ+X$           &  $Z(Q\overline Q)$ &   $ZQj+X$    & $ZQ\overline Q+X$ & $Z(Q\overline Q)j$  \\
\hline
 bottom& (8.23) 18.1&2.1&(2.19) $5.56^{+1.2}_{-0.9}\,^{+0.07}_{-0.05}\,^{+0.5}_{-0.5}$&(2.49) $3.74^{+0.45}_{-0.45}\,^{+0.12}_{-0.12}\,^{+0.15}_{-0.15}$&$0.63^{+0.26}_{-0.16}\,^{+0.06}_{-0.06}\,^{+0.05}_{-0.05}$\\
 charm & (11.3) 27.5&6.6&(3.21) $8.23^{+1.7}_{-1.4}\,^{+0.15}_{-0.26}\,^{+0.8}_{-0.8}$&(2.35) $3.47^{+0.44}_{-0.37}\,^{+0.0}_{-0.87}\,^{+0.14}_{-0.14}$&$2.08^{+0.85}_{-0.53}\,^{+0.22}_{-0.16}\,^{+0.15}_{-0.15}$\\

\hline
 all jets  & \multicolumn{2}{|c|}{(898) 1070} &  \multicolumn{3}{c|}{(163) $205^{+19}_{-19}\,^{+7}_{-2}\,^{+5}_{-5}$}\\
\hline
\end{tabular}
\end{center}
\label{tab:tev_incl}
\end{table}


\begin{table}[p]
\caption{Same as Table~\ref{tab:tev_incl}, except at the LHC
($\sqrt{s}=14$ TeV $pp$).  A jet lies in the range $p_T>15$ GeV
and $|\eta|<2.5$.}

\addtolength{\arraycolsep}{0.2cm}
\renewcommand{\arraystretch}{1.5}
\medskip
\begin{center}
\begin{tabular}[5]{|c|cc|ccc|}
\hline \hline \multicolumn{1}{|c|}{LHC} &
\multicolumn{5}{c|}{$\sigma$ (pb)}\\[1pt]
\hline
& \multicolumn{2}{|c|}{$Z+1$ jet $+X$} & \multicolumn{3}{c|}{$Z+2$ jets $+X$} \\
\cline{2-6}
           & $ZQ+X$        &  $Z(Q\overline Q)$ & $ZQj+X$   & $ZQ\overline Q+X$ & $Z(Q\overline Q)j$  \\
\hline
 bottom& (826) 1060&25&(353) $513^{+84}_{-58}\,^{+44}_{-35}\,^{+25}_{-25}$&(111) $153^{+20}_{-20}\,^{+2}_{-2}\,^{+9}_{-9}$&$24^{+10}_{-6}\,^{+0.3}_{-0.3}\,^{+1.2}_{-1.2}$\\
 charm & (989) 1430&50&(443) $746^{+110}_{-110}\,^{+0}_{-46}\,^{+45}_{-45}$&(90) $125^{+17}_{-17}\,^{+2}_{-2}\,^{+8}_{-8}$&$59^{+23}_{-15}\,^{+2}_{-2}\,^{+3}_{-3}$\\

\hline
 all jets  & \multicolumn{2}{|c|}{ (15300) 18400} &  \multicolumn{3}{c|}{(6090) $6650^{+470}_{-500}\,^{+170}_{-50}\,^{+240}_{-240}$}\\
\hline
\end{tabular}
\end{center}
\label{tab:lhc_incl}
\end{table}

\begin{figure}[t]
\begin{center}
\includegraphics[angle=90,width=14cm]{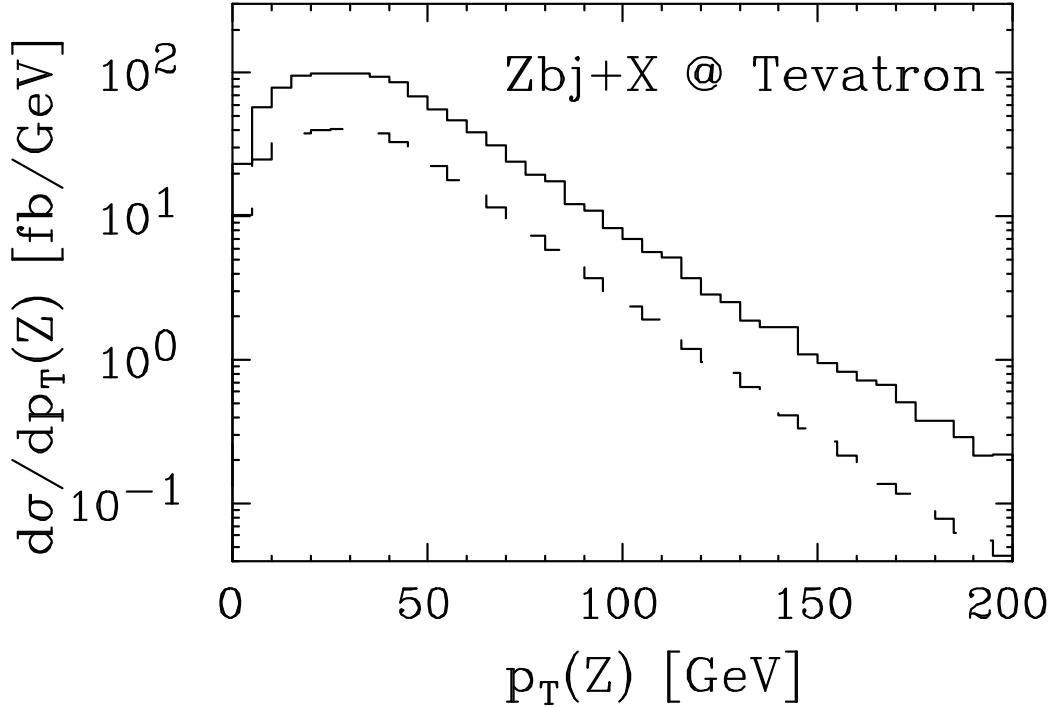}
\end{center}
\caption{Transverse-momentum distribution of the $Z$ boson in events
with at least two jets, one of which contains a bottom quark, at the
Tevatron ($\sqrt{s}=1.96$ TeV $p\bar p$).  The jets satisfy the
conditions $p_T>15$ GeV, $|\eta|<2$, and $\Delta R_{jj}>0.7$.  Both
LO (dashed) and NLO (solid) distributions are shown.}
\label{fig:pt_tev}
\end{figure}
\begin{figure}[t]
\begin{center}
\vspace*{0cm} \hspace*{0cm}
\includegraphics[angle=90,width=14cm]{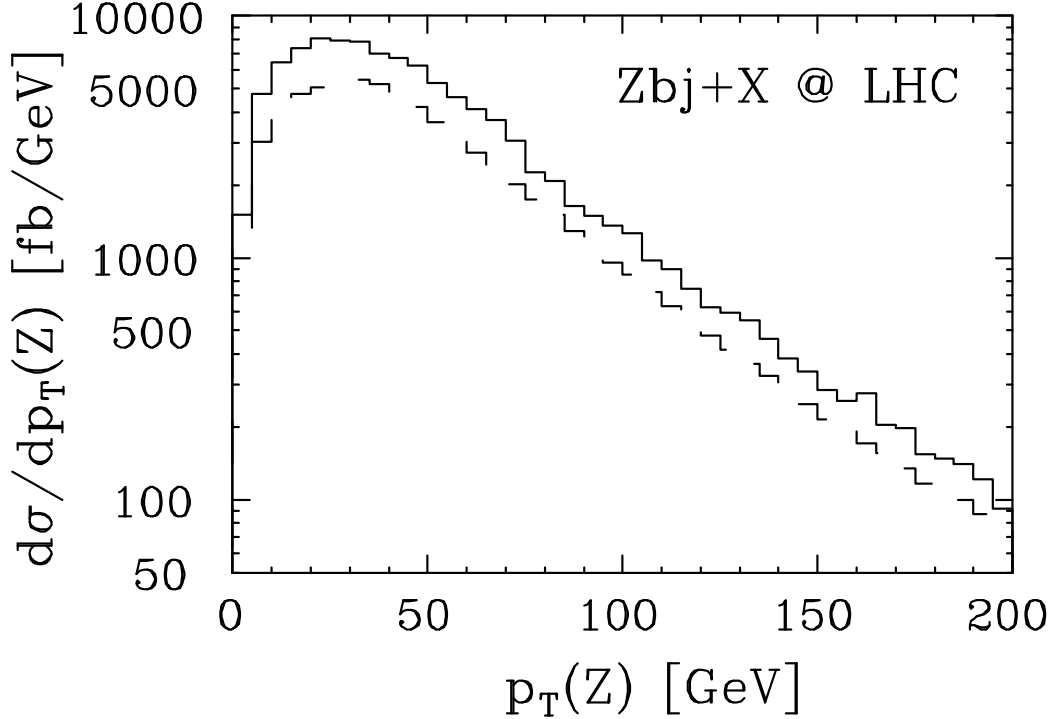}
\end{center}
\caption{Same as Fig.~\ref{fig:pt_tev}, but at the LHC
($\sqrt{s}=14$ TeV $pp$).  The jets satisfy the conditions $p_T>15$
GeV, $|\eta|<2.5$, and $\Delta R_{jj}>0.7$.} \label{fig:pt_lhc}
\end{figure}

We show in Figures~\ref{fig:pt_tev} and \ref{fig:pt_lhc} the
transverse-momentum spectrum of the $Z$ boson in events with at
least two jets, one of which contains a bottom quark, at the
Tevatron and the LHC. Both the LO and NLO distributions are shown.
The radiative corrections do not significantly change the shape of
the distribution, as is often the case when the quantity plotted
is not affected by the change in the kinematics due to additional
radiation.

\section{Conclusions}

In this paper we present a NLO calculation of the production of a
$Z$ boson plus two jets, one or more of which contains a heavy
quark. This greatly improves the accuracy with which this
fundamental background is known at the Tevatron and the LHC.  We
provide our results in both an exclusive (Tables~\ref{tab:tev_excl}
and \ref{tab:lhc_excl}) and an inclusive manner
(Tables~\ref{tab:tev_incl} and \ref{tab:lhc_incl}).  The NLO cross
section is significantly greater than the LO cross section for the
default scale choice $\mu_F=\mu_R=M_Z$.

Our calculation makes use of the heavy quarks present in the
proton sea, which are perturbatively calculable.  This makes the
LO calculation simpler, thus allowing a higher-order calculation
to be tractable.  We showed that the processes $Qq(g)\to ZQq(g)$
are a significant source of $ZQj$ events at the Tevatron, and the
dominant source at the LHC.

Alternatively, one could eschew the heavy quarks in the proton sea,
and regard the proton as containing only light quarks and gluons. In
that approach, the LO processes for $ZQj$ are $gq(g)\to ZQ\overline
Qq(g)$, where one of the heavy quarks is outside the acceptance of
the detector.  We give the results of this calculation in square
brackets in Table~\ref{tab:lo}, using CTEQL1 with $\mu_F=\mu_R=M_Z$.
The results are quite consistent with the LO results of the
heavy-quark approach at the Tevatron, but at the LHC they lie
somewhat below, though within a factor of two. In any case, the most
accurate cross sections are the NLO results presented in this paper,
based on the heavy-quark approach.

An important application of these results is to the search for the
Higgs boson via $q\bar q\to Zh$, followed by $h\to b\bar b$, where
the signal is $Z+2j$ with one or two $b$ tags \cite{Acosta:2005ga}.
We provide NLO results for the backgrounds with either one or two
heavy-quark jets. We have shown that the majority of such events
have only one heavy-quark jet. We see from Tables~\ref{tab:tev_incl}
and \ref{tab:lhc_incl} that $Zbj+X$ is nearly twice as big as
$Zb\bar b+X$ at the Tevatron, and more than three times as large at
the LHC. Similarly, $Zcj+X$ is about thrice $Zc\bar c+X$ at the
Tevatron, and about six times as large at the LHC.

\section*{Acknowledgments}

\indent\indent We are grateful for conversations with C.~Ciobanu,
T.~Junk, T.~Liss, S.~Lowette, B.~Mellado, K.~Pitts, P.~Venlaer,
and C.~Weiser. J.~C. and S.~W.~thank the Aspen Center for Physics for
hospitality. This work was supported in part by the
U.~S.~Department of Energy under contracts Nos.~DE-AC02-76CH03000
and DE-FG02-91ER40677.


\begin{titlepage}

\hspace*{\fill}\parbox[t]{5.5cm} {
\today} \vskip2cm
\begin{center}
{\Large \bf Erratum: Production of a $Z$ Boson and Two Jets \\
\bigskip  with One Heavy-Quark Tag} \\
\medskip
\bigskip\bigskip\bigskip\bigskip
{\large  {\bf J.~Campbell}$^1$,
         {\bf R.~K.~Ellis}$^2$,
         {\bf F.~Maltoni}$^3$,
     and {\bf S.~Willenbrock}$^4$} \\
\bigskip\bigskip\medskip
$^{1}$Department of Physics and Astronomy, University of Glasgow \\
Glasgow G12 8QQ, United Kingdom \\
\bigskip
$^{2}$Theoretical Physics Department, Fermi National Accelerator Laboratory \\
P.~O.~Box 500, Batavia, IL\ \ 60510 \\ \bigskip
$^{3}$Institut de Physique Th\'{e}orique and \\
Centre for Particle Physics and Phenomenology (CP3) \\
Universit\'{e} Catholique de Louvain\\[1mm]
Chemin du Cyclotron 2 \\
B-1348 Louvain-la-Neuve, Belgium \\
\bigskip
$^{4}$Department of Physics, University of Illinois at Urbana-Champaign \\
1110 West Green Street, Urbana, IL\ \ 61801 \\ \bigskip
\end{center}

\bigskip\bigskip\bigskip

\end{titlepage}

We recently presented a next-to-leading-order (NLO) calculation of
$Z+2$ jets, with one or more heavy-quarks ($Q=c,b)$, at hadron
colliders \cite{Campbell:2005zv}.  This process is a background to
numerous searches at hadron colliders, in particular the search for
the Higgs boson: it is the principal irreducible background to
$q\bar q\to Zh$, followed by $h\to b\bar b$, and a significant
reducible background to $gg\to h\to ZZ\to \ell^+\ell^-\ell^+\ell^-$
when both jets contain heavy quarks that decay semi-leptonically. We
neglected the heavy-quark mass throughout that calculation, except
when two heavy quarks were produced from a virtual gluon, $g^*\to
Q\bar Q$, in which case it is mandatory to keep the quark mass
nonzero when the heavy quarks are collinear. This occurs when two
heavy quarks are contained in the same jet. More generally, one
cannot neglect the quark mass if the invariant mass of the quark
pair coming from $g^*\to Q\bar Q$ is not large compared with the
quark mass.

In a subsequent NLO calculation of $W+2$ jets, with one or more $b$
quarks, we learned that this can also occur in another situation
\cite{Campbell:2006cu}, which we overlooked in
Ref.~\cite{Campbell:2005zv}. If both heavy quarks are at large
transverse momentum ($p_T$) compared to their mass, then the mass
may be safely neglected \cite{FebresCordero:2006sj}. However, there
are NLO processes in which one heavy quark is at high $p_T$ while
another is at low $p_T$ (and is missed by the detector), yet their
invariant mass is not large compared to their mass.  If this
heavy-quark pair comes from a virtual gluon splitting to a
heavy-quark pair, $g^*\to Q\bar Q$, then one cannot neglect the
quark mass. This can occur in the NLO processes $q\bar q\to ZQ\bar
Qg$ (Fig.~\ref{fig:ZQQg}) and $gq\to ZQ\bar Qq$
(Fig.~\ref{fig:ZQQj}), both of which contribute to $ZQj$ when one
heavy quark is missed.

\setcounter{figure}{0}

\begin{figure}[ht]
\begin{center}
\vspace*{0cm} \hspace*{0cm} \epsfxsize=5cm \epsfbox{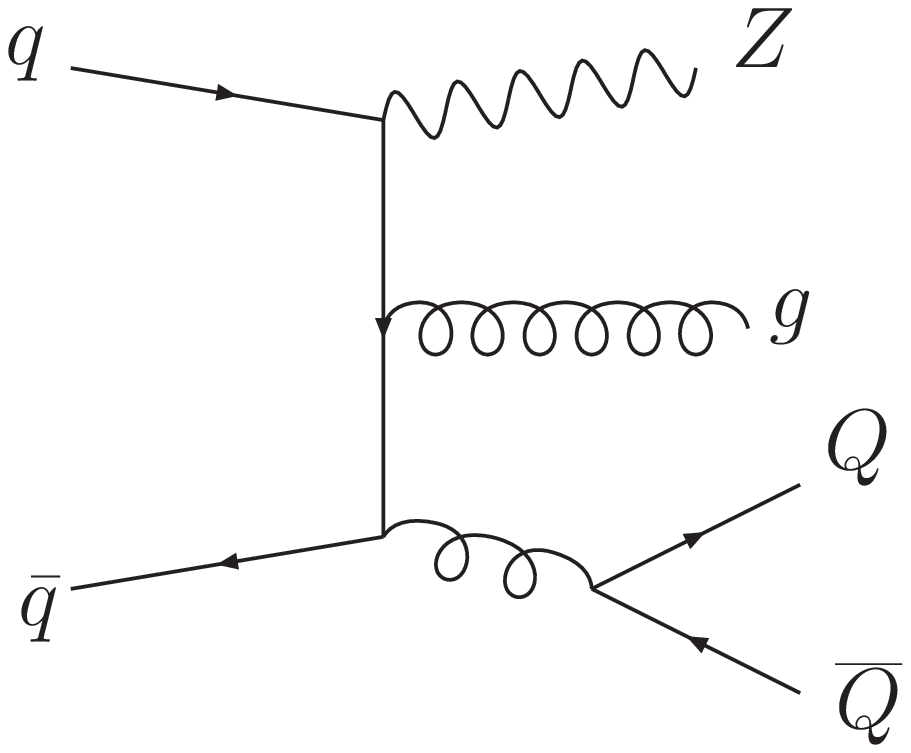}
\vspace*{-.8cm}
\end{center}
\caption{Diagram contributing to the NLO correction to the
associated production of a $Z$ boson and two high-$p_T$ jets, one or
more of which contains a heavy quark ($Q=c,b$).} \label{fig:ZQQg}
\end{figure}

\begin{figure}[ht]
\begin{center}
\vspace*{0cm} \hspace*{0cm} \epsfxsize=16cm \epsfbox{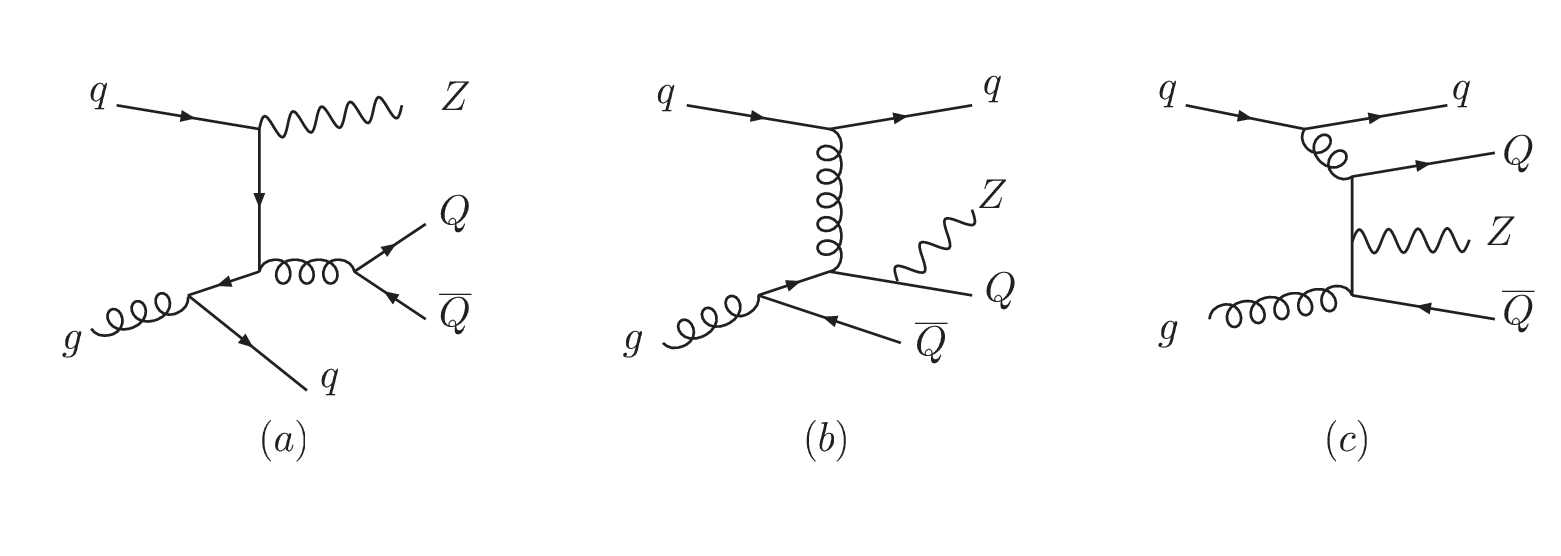}
\vspace*{-.8cm}
\end{center}
\caption{Diagrams contributing to the NLO correction to the
associated production of a $Z$ boson and two high-$p_T$ jets, one or
more of which contains a heavy quark ($Q=c,b$).} \label{fig:ZQQj}
\end{figure}

\begin{figure}[ht]
\begin{center}
\includegraphics[angle=90,width=14cm]{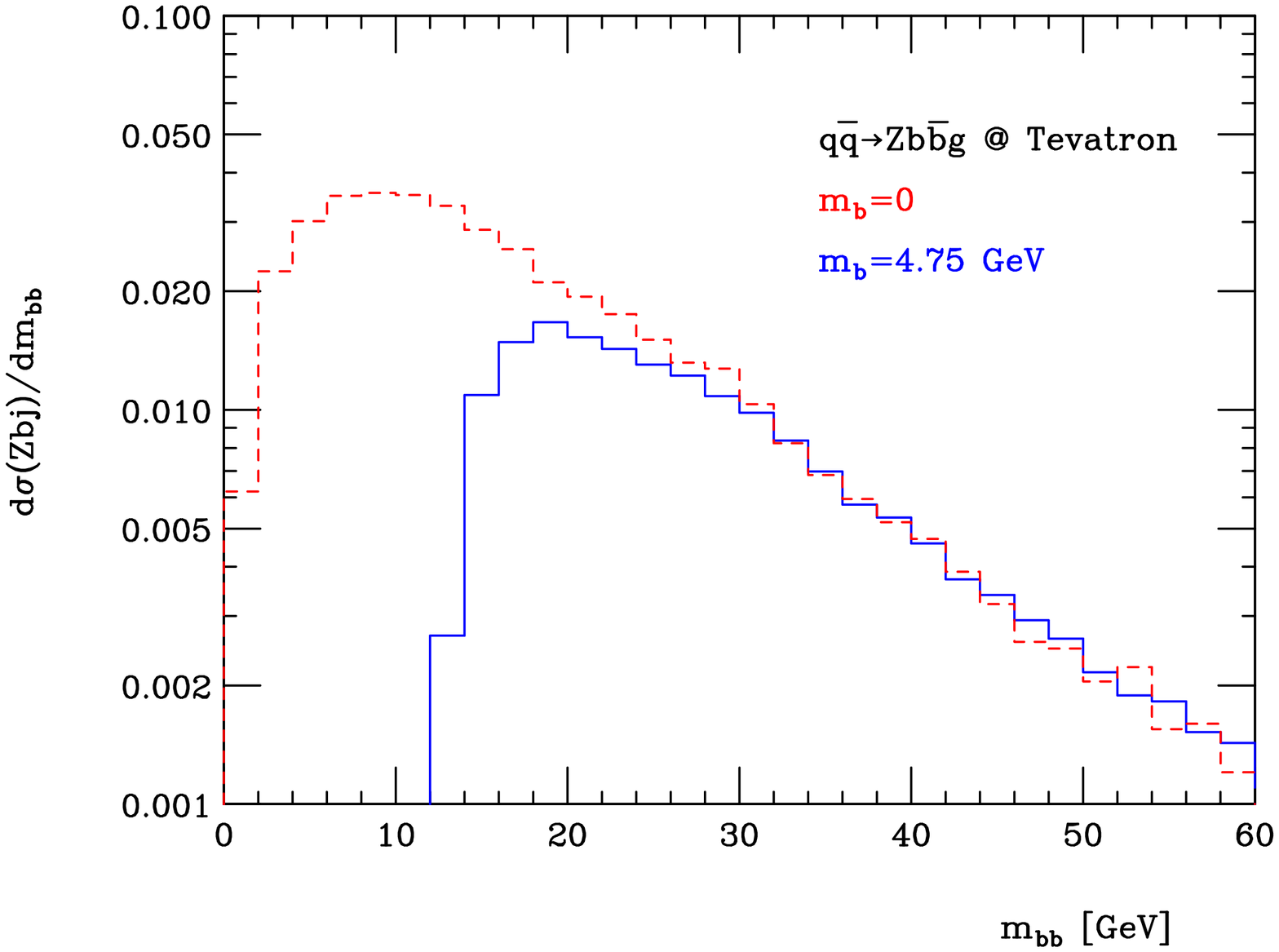}
\vspace*{-.8cm}
\end{center}
\caption{The differential cross section for $Z$ plus two jets, one
of which contains a $b$ quark, vs. the invariant mass of the $b$
quark and another $b$ quark that lies outside the acceptance of the
detector (taken to be $p_T>15$ GeV, $|\eta|<2$, $\Delta
R_{jj}>0.7$), at the Fermilab Tevatron ($\sqrt S=1.96$ TeV $p\bar
p$).  Only the contribution from the subprocess $q\bar q\to Zb\bar
bg$ is shown. The (solid, blue) curve includes the $b$ quark mass,
while the (dashed, red) curve does not.} \label{fig:tev-b}
\end{figure}

To demonstrate this, we show in Fig.~\ref{fig:tev-b} the
invariant-mass distribution of the heavy-quark pair from $q\bar q\to
Zb\bar bg$ when it contributes to the final state $Zbj$, with the
other heavy quark outside the acceptance of the detector (taken to
be $p_T>15$ GeV, $|\eta|<2$, $\Delta R_{jj}>0.7$), at the Fermilab
Tevatron ($\sqrt S=1.96$ TeV $p\bar p$).  There are two curves, one
with (solid, blue) and one without (dashed, red) the quark mass
included. It is clear that the quark mass has a large influence on
the cross section (the area under the curves).

In the case of the process $gq\to ZQ\bar Qq$, which involves the
splitting $g\to Q\bar Q$ in the initial state
[Fig.~\ref{fig:ZQQj}(b)] as well as the final state
[Fig.~\ref{fig:ZQQj}(a)], this requires that we abandon the dipole
subtraction method \cite{Catani:1996vz} in favor of subtracting the
mass singularity via the truncated $Q$ distribution function
\cite{Aivazis:1993pi,Collins:1998rz}
\begin{equation}
\tilde Q(x,\mu) =
\frac{\alpha_S(\mu)}{2\pi}\ln\left(\frac{\mu^2}{m_Q^2}\right)\int_x^1\frac{dy}{y}P_{qg}\left(\frac{x}{y}\right)g(y,\mu)
\end{equation}
where $P_{qg}(z)={1\over 2}[z^2+(1-z)^2]$ is the DGLAP splitting
function.  The counterterm is constructed by calculating $\tilde
Qq\to ZQq$; this cancels the initial-state logarithmic dependence on
the heavy-quark mass in $gq\to ZQ\bar Qq$, and yields a cross
section in the $\overline{\rm MS}$ factorization scheme.  This
technique was developed in Ref.~\cite{Campbell:2006cu}.

We have revisited our NLO calculation of $Z+2$ jets, with one or
more heavy quarks, and corrected this omission.  Specifically, we
recalculated the NLO processes\footnote{It is not necessary to
recalculate $gg\to ZQ\overline Qg$, since there is no contribution
from a virtual gluon splitting to a heavy-quark pair, $g^*\to Q\bar
Q$.  We recalculated it nevertheless, and found that the effect of
the heavy-quark mass is about $-6\%$ at the Tevatron and $-12\%$ at
the LHC.  However, since this subprocess makes a relatively small
contribution to the cross section, this change is negligible.}
\begin{itemize}
\item $q\bar q\to ZQ\overline Qg$ at tree level [Fig.~\ref{fig:ZQQg}]
\item $gq\to ZQ\overline Qq$ at tree level [Fig.~\ref{fig:ZQQj}]
\end{itemize}
using a finite heavy-quark mass throughout ($m_c = 1.4$ GeV, $m_b =
4.75$ GeV).

The results for the exclusive and inclusive cross sections at the
Fermilab Tevatron and the CERN Large Hadron Collider (LHC) are
presented in Tables~II-V.  These should replace the corresponding
tables in Ref.~\cite{Campbell:2005zv}.  While there are small
changes throughout the tables, the only significant change is in the
NLO cross section for $Zbj$ at the Tevatron in
Table~\ref{tab:tev_excl} (as well as the inclusive cross section
$Zbj+X$ in Table~\ref{tab:tev_incl}), which was reduced by 16\% by
including the $b$-quark mass (5.23 pb $\to$ 4.37 pb).  The dominant
effect of the $b$-quark mass was on the subprocess $q\bar q\to
ZQ\bar Qg$ [Fig.~\ref{fig:ZQQg}], which was reduced by more than
50\% (1.16 pb $\to$ 0.50 pb).  A similar percentage reduction occurs
at the LHC (9 pb $\to$ 4 pb), but this subprocess is so small at
that machine that it hardly affects the total result for $Zbj$.

We raised the $p_T$ threshold on the acceptance for $b$ jets to
greater than 15 GeV, and found that the effect of the $b$ mass
became less. This is consistent with our findings that the effect of
the $c$ mass is smaller than that of the $b$ mass for the same $p_T$
threshold.

\setcounter{table}{1}


\begin{table}[t]
\caption{Cross sections (pb) for $Z$-boson plus two (or more) jets,
one or more of which contains a heavy quark, at the Tevatron
($\sqrt{s}=1.96$ TeV $p\bar p$). A jet lies in the range $p_T>15$
GeV and $|\eta|<2$. Two final-state partons are merged into a single
jet if $\Delta R_{jj}<0.7$. No branching ratios or tagging
efficiencies are included. Numbers in parenthesis are leading-order
results. The labels on the columns have the following meaning:
$ZQj=$ exactly two jets, one of which contains a heavy quark;
$ZQ\overline Q=$ exactly two jets, both of which contain a heavy
quark; $Z(Q\overline Q)j=$ exactly two jets, one of which contains a
heavy-quark pair; $ZQ\overline Qj=$ exactly three jets, two of which
contain a heavy quark; $ZQjj=$ exactly three jets, one of which
contains a heavy quark.  For the last set of processes, the labels
mean: $Zjj=$ exactly two jets, including heavy quarks; $Zjjj=$
exactly three jets, including heavy quarks. For $ZQj$ and
$ZQ\overline Q$, both the leading-order (in parentheses) and
next-to-leading-order cross sections are given. The CTEQ6M parton
distribution functions are used throughout, except for the LO cross
sections in parentheses, where CTEQ6L1 is used
\cite{Pumplin:2002vw}. The factorization and renormalization scales
are chosen as $\mu_F=\mu_R=M_Z$.}

\addtolength{\arraycolsep}{0.2cm}
\renewcommand{\arraystretch}{1.5}
\medskip

\begin{center}
\begin{tabular}[5]{|c|ccccc|}
\hline \hline \multicolumn{1}{|c|}{Tevatron} &
\multicolumn{5}{c|}{$\sigma$ (pb)}\\[1pt]
\hline
         &  $ZQj$  & $ZQ\overline Q$ & $Z(Q\overline Q)j$  &   $ZQ\overline Qj$  &  $ZQjj$ \\
\hline
bottom   & (2.18) 4.37  & (2.47) 3.07  &  0.634 &  0.672 & 0.326 \\
charm    & (3.20) 7.35  & (2.34) 2.75  &  2.00  &  0.621 & 0.495 \\
\hline & \multicolumn{3}{|c}{$Zjj$} & \multicolumn{2}{c|}{$Zjjj$} \\
\hline $Z+$jets & \multicolumn{3}{|c}{(163) 182} & \multicolumn{2}{c|}{22.9} \\
\hline
\end{tabular}
\end{center}

\label{tab:tev_excl}
\end{table}


\begin{table}[t]
\caption{Same as Table~\ref{tab:tev_excl}, except at the LHC
($\sqrt{s}=14$ TeV $pp$).  A jet lies in the range $p_T>15$ GeV and
$|\eta|<2.5$.  (This table is nearly identical to Table~III of
Ref.~\cite{Campbell:2005zv}.)}

\addtolength{\arraycolsep}{0.2cm}
\renewcommand{\arraystretch}{1.5}
\medskip
\begin{center}
\begin{tabular}[5]{|c|ccccc|}
\hline \hline \multicolumn{1}{|c|}{LHC} &
\multicolumn{5}{c|}{$\sigma$ (pb)}\\[1pt]
\hline
         &  $ZQj$  & $ZQ\overline Q$ & $Z(Q\overline Q)j$  &   $ZQ\overline Qj$  &  $ZQjj$ \\
\hline
bottom   & (352) 418 & (109)  92.1  &  23.5 &  60.8 & 92.1 \\
charm    & (443) 624 & (87) 75.2  &  58.6 &  49.3 & 123 \\
\hline & \multicolumn{3}{|c}{$Zjj$} & \multicolumn{2}{c|}{$Zjjj$} \\
\hline $Z+$jets & \multicolumn{3}{|c}{(6090) 4840} & \multicolumn{2}{c|}{1810} \\
\hline
\end{tabular}
\end{center}

\label{tab:lhc_excl}
\end{table}



\begin{table}[p]
\caption{Inclusive cross sections (pb) for $Z$ boson plus one or two
jets, one or two of which contains a heavy quark, at the Tevatron
($\sqrt{s}=1.96$ TeV $p\bar p$). A jet lies in the range $p_T>15$
GeV and $|\eta|<2$. Two final-state partons are merged into a single
jet if $\Delta R_{jj}<0.7$. No branching ratios or tagging
efficiencies are included. Numbers in parenthesis are leading-order
results. The labels on the columns have the following meaning:
$ZQ+X=$ at least one jet, at least one of which contains a heavy
quark; $Z(Q\overline Q)=$ one jet which contains a heavy-quark pair;
$ZQj+X=$ at least two jets, one of which contains a heavy quark;
$ZQ\overline Q+X=$ at least two jets, two of which contain a heavy
quark; $Z(Q\overline Q)j=$ two jets, one of which contains a
heavy-quark pair.  The last row gives the inclusive cross section
for jets containing both light and heavy partons. The CTEQ6M parton
distribution functions are used throughout, except for the LO cross
sections in parentheses, where CTEQ6L1 is used
\cite{Pumplin:2002vw}.  The factorization and renormalization scales
are chosen as $\mu_F=\mu_R=M_Z$.  The uncertainties are from the
variation of the renormalization scale, the factorization scale, and
the parton distribution functions, respectively.}

\addtolength{\arraycolsep}{0.2cm}
\renewcommand{\arraystretch}{1.5}
\medskip
\begin{center}
\hspace*{-1.2cm}
\begin{tabular}[5]{|c|cc|ccc|}
\hline \hline \multicolumn{1}{|c|}{Tevatron} &
\multicolumn{5}{c|}{$\sigma$ (pb)}\\[1pt]
\hline
& \multicolumn{2}{|c|}{$Z+1$ jet $+X$} & \multicolumn{3}{c|}{$Z+2$ jets $+X$} \\
\cline{2-6}
           & $ZQ+X$           &  $Z(Q\overline Q)$ &   $ZQj+X$    & $ZQ\overline Q+X$ & $Z(Q\overline Q)j$  \\
\hline
 bottom& (8.23) 18.1&2.1&(2.18) $4.70^{+1.2}_{-0.9}\,^{+0.07}_{-0.05}\,^{+0.5}_{-0.5}$&(2.47) $3.74^{+0.45}_{-0.45}\,^{+0.12}_{-0.12}\,^{+0.15}_{-0.15}$&$0.63^{+0.26}_{-0.16}\,^{+0.06}_{-0.06}\,^{+0.05}_{-0.05}$\\
 charm & (11.3) 27.5&6.6&(3.20) $7.85^{+1.7}_{-1.4}\,^{+0.15}_{-0.26}\,^{+0.8}_{-0.8}$&(2.34) $3.37^{+0.44}_{-0.37}\,^{+0.0}_{-0.87}\,^{+0.14}_{-0.14}$&$2.00^{+0.85}_{-0.53}\,^{+0.22}_{-0.16}\,^{+0.15}_{-0.15}$\\

\hline
 all jets  & \multicolumn{2}{|c|}{(898) 1070} &  \multicolumn{3}{c|}{(163) $205^{+19}_{-19}\,^{+7}_{-2}\,^{+5}_{-5}$}\\
\hline
\end{tabular}
\end{center}
\label{tab:tev_incl}
\end{table}


\begin{table}[p]
\caption{Same as Table~\ref{tab:tev_incl}, except at the LHC
($\sqrt{s}=14$ TeV $pp$).  A jet lies in the range $p_T>15$ GeV and
$|\eta|<2.5$.  (This table is nearly identical to Table~V of
Ref.~\cite{Campbell:2005zv}.)}

\addtolength{\arraycolsep}{0.2cm}
\renewcommand{\arraystretch}{1.5}
\medskip
\begin{center}
\begin{tabular}[5]{|c|cc|ccc|}
\hline \hline \multicolumn{1}{|c|}{LHC} &
\multicolumn{5}{c|}{$\sigma$ (pb)}\\[1pt]
\hline
& \multicolumn{2}{|c|}{$Z+1$ jet $+X$} & \multicolumn{3}{c|}{$Z+2$ jets $+X$} \\
\cline{2-6}
           & $ZQ+X$        &  $Z(Q\overline Q)$ & $ZQj+X$   & $ZQ\overline Q+X$ & $Z(Q\overline Q)j$  \\
\hline
 bottom& (826) 1060&25&(352) $510^{+84}_{-58}\,^{+44}_{-35}\,^{+25}_{-25}$&(109) $153^{+20}_{-20}\,^{+2}_{-2}\,^{+9}_{-9}$&$24^{+10}_{-6}\,^{+0.3}_{-0.3}\,^{+1.2}_{-1.2}$\\
 charm & (989) 1430&50&(443) $747^{+110}_{-110}\,^{+0}_{-46}\,^{+45}_{-45}$&(87) $125^{+17}_{-17}\,^{+2}_{-2}\,^{+8}_{-8}$&$59^{+23}_{-15}\,^{+2}_{-2}\,^{+3}_{-3}$\\

\hline
 all jets  & \multicolumn{2}{|c|}{ (15300) 18400} &  \multicolumn{3}{c|}{(6090) $6650^{+470}_{-500}\,^{+170}_{-50}\,^{+240}_{-240}$}\\
\hline
\end{tabular}
\end{center}
\label{tab:lhc_incl}
\end{table}


\section*{Acknowledgments}

\indent\indent We are grateful for conversations with Gavin Salam.
This work was supported in part by the U.~S.~Department of Energy
under contracts Nos.~DE-AC02-76CH03000 and DE-FG02-91ER40677.


\end{document}